\documentclass[journal]{IEEEtran}

\usepackage[table,xcdraw]{xcolor}
\usepackage[pdftex]{graphicx}
\usepackage[
singlelinecheck=false
]{caption}
\usepackage[cmex10]{amsmath}

\usepackage{amssymb}
\usepackage{colortbl}

\usepackage{array}
\usepackage{mdwmath}
\usepackage{mdwtab}
\usepackage{eqparbox}
\usepackage{url}
\usepackage{algorithm}
\usepackage{algorithmic}
\usepackage{multirow}
\let\labelindent\relax
\usepackage{enumitem}
\usepackage{cancel}
\usepackage{booktabs}
\usepackage [
    n,
    advantage,
    operators,
    sets,
    adversary,
    landau,
    probability,
    notions,
    logic,
    ff,
    mm,
    primitives,
    events,
    complexity,
    asymptotics,
    keys
]{cryptocode}

\usepackage{calc}

\begin{document}
\title{SigK: rPPG-based key agreement scheme\\ for mHealth}

\author{\IEEEauthorblockN{Yanping Wang\IEEEauthorrefmark{2},
Lin Li\IEEEauthorrefmark{1},
Leo Yu Zhang\IEEEauthorrefmark{4},
Jianfeng Wang\IEEEauthorrefmark{5},
Chao Chen\IEEEauthorrefmark{1},
Kok-Leong Ong\IEEEauthorrefmark{1},
Jun Zhang\IEEEauthorrefmark{3}
}

\IEEEauthorblockA{\IEEEauthorrefmark{2}Xihua University, China},
\IEEEauthorblockA{\IEEEauthorrefmark{1}RMIT University, Australia}

\IEEEauthorblockA{\IEEEauthorrefmark{3}Swinburne University of Technology, Australia}

\IEEEauthorblockA{\IEEEauthorrefmark{4}Griffith University, Australia},
\IEEEauthorblockA{\IEEEauthorrefmark{5}Xidian University, China}

}

\maketitle

\begin{abstract}

As mobile health advances, the use of biomedical sensors for disease monitoring, diagnosis, and treatment is increasing. These sensors often need to be connected through wireless communication technologies to process physiological information from various body parts effectively. Since the data transmitted in these devices usually contain sensitive personal medical information, it is crucial to establish secure communication between devices. In recent studies, the inter-pulse intervals between heartbeats are often used as the key source. They can be extracted from physiological signals such as photoplethysmogram (PPG) or electrocardiogram (ECG), allowing key agreement schemes without pre-key distribution. However, traditional solutions are based on contact measurement devices {(e.g., wristbands, chest straps), which limit their applicability in contactless healthcare scenarios.} In this paper, we propose a remote PPG-based key agreement scheme to facilitate secure communication between contactless medical devices.
{We leverage medical devices equipped with cameras to capture remote PPG signals from human faces, and then utilize the signal's power spectral density and peak-peak interval to establish a key agreement with signals synchronously acquired in the Body Area Network (BAN).}
We have carried out performance evaluations on three publicly available datasets to demonstrate the feasibility of our proposed scheme. In addition, we have also presented a comprehensive proof of the security of our proposed scheme.

\end{abstract}


\IEEEpeerreviewmaketitle

\section{Introduction}
\label{intro}
In mobile health (mHealth), the adoption of biosensors on individuals has become increasingly prevalent. Devices such as smartwatches and implantable medical devices enable real-time monitoring of various physiological parameters, including heart rate, blood glucose levels, and physical activity~\cite{li2025smart}.
The information obtained from these biosensors provides valuable insights for healthcare providers. It facilitates remote monitoring by healthcare professionals, allowing for prompt response to changes in a patient's condition and delivering timely interventions \cite{tan2024systematic,wang2024trustmis}.

\begin{figure}[!t]
  \centering
  \includegraphics[width=\linewidth ]{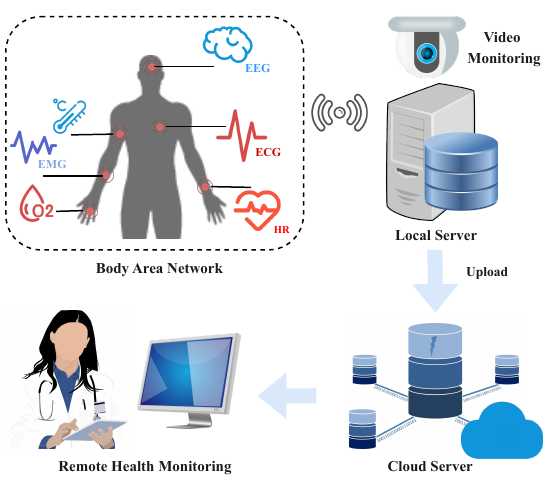}
  \caption{An overview of mHealth workflow. Body Area Networks transmit physiological signals to a local service for storage in the cloud. The mHealth team then uses this data to monitor the user's health status.}
  \label{workflowr}
\end{figure}

In addition to the clinical benefits, mHealth also brings new challenges.
Since mHealth technology handles sensitive personal health data, ensuring wireless communication security for biosensors is crucial to protect patient privacy and prevent data breaches \cite{salehzadeh2024wearable}. Any data breaches or unauthorized access to sensitive medical information could have serious consequences for patients \cite{wang2022towards}.
Encryption technology is commonly used to ensure the security of medical data and authorized access.
Traditional symmetric encryption requires both communication parties to share a key through a key distribution scheme before encryption occurs.
However, key distribution technologies like random key pre-distribution and polynomial pool-based methods require pre-deployment.
They can cause delays during network initialization or adjustments, as well as require reconfiguration whenever sensors are added or changed.
It also brings significant energy consumption in the communication between the biosensor and the monitoring device or server \cite{ai2025improved}.
Moreover, biosensors are often limited by computational resources and energy sources, making it challenging to implement traditional encryption solutions \cite{wang2024towards}. As such, alternative encryption solutions are required.

An increasing number of studies have revealed that physiological signals can be used to effectively establish key agreement schemes uniquely and securely \cite{zhang9483626,guglielmi2021information}.
In particular, the inter-pulse interval (IPI) of the heartbeat, which refers to the time difference between two consecutive heartbeats, has been identified as a valuable source for extracting encryption keys~\cite{zhao9863646}.
IPI can be extracted from various physiological signals associated with cardiac activity, such as photoplethysmogram (PPG) and electrocardiogram (ECG).
Taking measurements on different body parts, such as through a wristband or a chest strap, makes it possible to avoid the need for pre-key distribution.
Additionally, these signals are unique to each individual and are difficult to replicate, making them a secure alternative to traditional key agreement schemes \cite{tang8388272}.

Despite these advantages, most existing studies rely on contact-based biosensors in body area networks (BAN), as shown in Fig.~\ref{workflowr}. However, contactless biosensors such as video surveillance equipment also play an essential role for mHealth \cite{casalino2022evaluating}, as it allows for remote monitoring and evaluation of patients. It enables healthcare professionals to remotely evaluate patients' conditions and provide treatment, even in the absence of physical presence. In addition, it can also be used to monitor the progress of patients recovering from surgery or an illness at home \cite{hoogland2019feasibility}.
Nevertheless, existing schemes are ill-equipped for this contactless paradigm, since they mainly assume that both parties obtain physiological signals through contact-based sensing modalities. Therefore, there is an urgent need to develop a key agreement mechanism that can bridge contact-based biosensors and contactless devices.

Recent advances in remote photoplethysmography (rPPG) offer a feasible way to address this challenge. Specifically, recent studies~\cite{shao2025remote,zhao2026phase,lee2023lstc} have showcased the feasibility of using videos of faces to extract heartbeat-related physiological signals, known as remote photoplethysmography (rPPG). This advancement creates new opportunities to extend heartbeat-based key agreements beyond contact biosensors to camera-enabled devices, thereby providing a promising foundation for secure, contactless, and versatile mHealth applications.

Motivated by these observations, we propose an rPPG-based key agreement scheme for secure mHealth applications. By exploiting the physiological consistency between PPG signals, the proposed scheme enables contact-based biosensors and camera-enabled devices to establish a secure session key without relying on pre-distributed secrets. The main contributions of this paper are summarized as follows:

\begin{itemize}

  \item We introduce a contactless key generation mechanism for mHealth. Unlike existing methods that derive key materials mainly from contact-based biosensors, our scheme further leverages rPPG to extract physiological signals, allowing camera-enabled terminals to obtain physiological key materials in a contactless manner.

  \item We propose a novel rPPG-based key agreement scheme for mHealth, named SigK, to fortify the communication between mHealth biosensors and  camera-enabled terminals. By enabling parties to establish a shared session key from physiological signals, SigK protects the confidentiality, integrity, and authenticity of transmitted data.

  \item  We provide a comprehensive security and performance evaluation of SigK. The security analysis shows that SigK can resist various security attacks, while experiments demonstrate its ability to generate consistent key materials with practical key generation rates.

\end{itemize}

This paper is structured as follows: We review related work in Section~\ref{relatedwork}, including the physiological signal-based key agreement scheme. Section~\ref{prelim} introduces the preliminaries. Section~\ref{method} describes our proposed rPPG-based key agreement scheme in mHealth.
We prove the security of the proposed scheme in Section~\ref{securitysec}. Then,
we comprehensively evaluate our proposed solution in Section~\ref{evaluations}, discuss its limitations in Section~\ref{discussion}, and conclude the work in Section~\ref{conclusion}.

\section{Related Works}
\label{relatedwork}
To address the issue of secure communication in mHealth, a physiological signal-based approach for protecting implanted biosensor communication was first proposed in 2003 \cite{cherukuri1240399}.
Then, Poon et al.~\cite{Poon1632652} employ human physiological signals to create a unique Entity Identifier (EI) for each network node, allowing identification and secure access control to the key.

Some physiological features, captured from multiple locations on the human body, are utilized to generate EI, enabling secure key distribution without the need for complex key agreement and exchange protocol \cite{Chunqiang6567031}. For instance, the inter-pulse interval (IPI) and heart rate variability (HRV) are considered appropriate physiological parameters for generating EIs \cite{seepers7312400}. Specifically, the sender and receiver use PPG/ECG signals to extract the IPI, which is then encoded as binary to encrypt the message. The sender transmits the encrypted message, and the receiver uses its PPG/ECG signal to recover the key for decryption.

The possibility of desynchronizing IPIs primarily arises from delays in cardiovascular pulse propagation and signal transmission. Synchronizing signal measurement between devices is a challenge.
Traditionally, the differences caused by desynchronization can be treated as communication errors, and implementing an error tolerance mechanism can reduce its impact \cite{Bao4534350}. Although error correction codes can correct for some discrepancies in the feature vector, they cannot address the reordering of the vector or the presence of additional features \cite{venkatasubramanian5352222}.
Additionally, the trends of time-domain components from different body parts (e.g., two fingers) are alike. However, there are disparities in their values.
Conversely, the frequency components of PPG signals taken from any part of the body exhibit consistent values at a specific moment. Compared to time-domain-based methods, a frequency-domain-based approach is less sensitive to signal quality \cite{venkatasubramanian2008plethysmogram}.

In addition to generating EI, physiological signals are also employed to facilitate key agreement. The physiological value-based scheme (PVS) employs biometric features from the deployment environment, such as IPI, to obfuscate the key through an XOR operation \cite{krishna1777410}. The receiver then uses its own version of the biometric feature to unmask the key and verify or decrypt the received data. To address potential errors in communication, they use a fuzzy-vault scheme compensation mechanism to account for differences in the sender's and receiver's features \cite{venkatasubramanian5352222}. In particular,
Choi et al.~\cite{choi2017e2pka} utilized regression analysis instead of the fuzzy vaulting scheme, resulting in a reduction of at least 90\% in the power consumption of the sensors. H2K \cite{zhang2021h2k} synchronized time differences by indexing each IPI with sequence numbers and used information reconciliation to correct any key mismatches between the sender and the receiver.

However, all the above approaches aim to secure communication between contact biosensors in mhealth applications. The data acquired by the sensor is usually transmitted to a base station, smartphone, or proprietary dongle (local terminals) before being transmitted to the cloud using an asymmetric encryption scheme. All communication links are vulnerable to potential security breaches \cite{fan2026what}. Protecting communications between sensors and the local terminal is critical in practical applications. Any security breach or compromise at this level can result in significant consequences for the privacy and security of the transmitted health data, which is often sensitive and personal in nature \cite{chen2023fedward,long2024fedcl}.

\section{Preliminaries}
\label{prelim}
In this section, we briefly recall the primary techniques employed within our scheme and the associated cryptographic assumptions used in our constructions. The key notations involved throughout this paper are summarized in Table \ref{tab0}.

\begin{table}[!htbp]
\caption{{Key notations in this paper.}}
\footnotesize
\label{tab0}
\begin{tabular}{ll}
\toprule
\textbf{Notation}                      & \textbf{Description} \\ \midrule
  $ID_{i}$                             & The unique identifier of entity $i$\\
  $\lambda$                            & Security parameter. \\
  $p$                                  & A large prime number. \\
  $\ell$                                  & Bit length of the password and of the hash outputs. \\
  $\mathbb{G}$                         & An additive group with order $p$. \\
  $\mathbb{Z}_p$                       & The set of integers modulo $p$, i.e., $\{0,\dots,p-1\}$.  \\
  $P$                                  & A generator of $\mathbb{G}$. \\
  $H$                                  & Cryptographic hash function, from $\{0,1\}^{\ast}$ to $\mathbb{G}$. \\
  $H_1, H_2$                      & Cryptographic hash functions, from $\{0,1\}^{\ast}$ to $\{0,1\}^{\ell}$.   \\
  $ssk$                                & The negotiated session key. \\
  AKA                                  & Authenticated key agreement scheme.\\
  $\mathcal{A}$                        & A probabilistic polynomial-time (PPT) adversary. \\
  $\mathsf{Adv}^{AKA}(\mathcal{A})$      & Advantage of $\mathcal{A}$ in breaking AKA security. \\ \bottomrule
\end{tabular}
\end{table}

{
\subsection{rPPG Method}

  \textbf{CHROM}~\cite{de2013robust} is built upon the skin's optical reflection. It posits that the light reflecting off the skin is a combination of diffuse and specular reflections. The PPG signal can be extracted from the diffuse reflective part. CHROM removes the specular reflection component through the utilization of chrominance.
Let $X$ and $Y$ represent two orthogonal axes, and $R$, $G$, and $B$ represent the three color channels in the video,
the rPPG signal value $S$ can be calculated as follows:
\begin{equation}
  X = 3 \times R - 2 \times G \notag,
\end{equation}
\begin{equation}
  Y = (1.5\times R)+G-(1.5 \times B)\notag,
\end{equation}
\begin{equation}
  \alpha =\sigma (X)/\sigma (Y)\notag,
\end{equation}
\begin{equation}
  S = X - \alpha \times Y,
\end{equation}
 where $\sigma$ represents the method of obtaining the standard deviation.

  \textbf{GREEN}~\cite{verkruysse2008remote} extracts the green channel directly from the RGB video, capitalizing on its favourable tolerance to motion artifacts. Thus, PPG signals $S$ can be represented as:
  \begin{equation}
  S = G.
\end{equation}

\textbf{PBV}~\cite{de2014improved} employs the distinctive variations in blood volume at different wavelengths to extract PPG signals. The PBV vector can be obtained via the following equation:
\begin{equation}
S_{pbv} = \frac { \begin{bmatrix} \sigma (R), \sigma (G), \sigma (B) \end{bmatrix} }{\sqrt {\sigma ^{2}(R)+\sigma ^{2}(G)+\sigma ^{2}(B)}}.
\end{equation}

 \textbf{LGI}~\cite{Pilz8575328} introduces local group invariance features to improve heart rate estimation from facial videos, even under challenging conditions such as motion artifacts and illumination variations. Specifically, the method applies singular value decomposition (SVD) to the temporally normalized color signal $C_{tn}(t)$, given by $C_{tn}(t) = U\Sigma V^{T}$.
Based on this decomposition, a projection plane $\mathbf{Q}$ is constructed as:
\begin{equation}
\mathbf{Q} = I - UU^{T},
\end{equation}
where $I$ denotes the identity matrix. The rPPG signal $S$ is then obtained by projecting $C_{tn}(t)$ onto $\mathbf{Q}$:
\begin{equation}
S = \mathbf{Q} \cdot C_{tn}(t).
\end{equation}

\textbf{POS}~\cite{wang2016algorithmic} is comparable to CHROM, as both methods are grounded in the skin-reflection model. The key idea of POS is to extract the rPPG signal by projecting temporally normalized RGB signals onto a plane orthogonal to the skin-tone direction~\cite{wang2016algorithmic}. Specifically, for the $t$-th frame, the spatially averaged RGB signal over the skin is denoted as $C(t) = [R(t), G(t), B(t)]^T$. This spatial averaging step can reduce camera quantization errors.  A sliding window is first applied to perform temporal normalization of the RGB signals:
\begin{equation*}
 C_w(t) = \sum_{i=t}^{t+w-1}C(i),
\end{equation*}
\begin{equation}
 C_{tn}(t) = \frac{ C_w(t)}{\mu ( C_w(t))},
\end{equation}
where $w$ denotes the time window, and $t$ indicates the $t$-th frame.
Then, the temporally normalized signal is projected onto the plane $\mathbf{Q}$:
\begin{equation}
\mathbf{Q} = \begin{pmatrix}0 & 1&-1\\ -2 & 1&1\end{pmatrix}.
\end{equation}

Finally, the rPPG signal $S$ is obtained as:
\begin{equation*}
X, Y = \mathbf{Q} \cdot C_{tn}(t),
\end{equation*}
\begin{equation}
  S = X + \alpha \cdot Y,
\end{equation}
where $\alpha=\sigma(X)/\sigma(Y)$.

}

{

\subsection{Assumptions}

The security of our scheme is based on standard computational assumptions in cryptography. We outline the key assumptions used in our security analysis below.

\textit{Assumption 1. Computational Diffie-Hellman (CDH) Assumption:} Let $\mathbb{G}$ be an additive group of prime order $p$. The CDH assumption states that for given input $\{P\in\mathbb{G}, a\cdot P, b\cdot P\} = \{P, X, Y\}$ of unknown $a, b\in \mathbb{Z}_p$, it is hard to output the $a\cdot b\cdot P =\text{CDH}(X, Y) $, where $\mathbb{Z}_p=[0,p-1]$.

\textit{Assumption 2. Password-Based Chosen-Basis Computational Diffie-Hellman (PCCDH) Assumption~\cite{alnahawi2024nice}:}
Let $\mathbb{G}$ be an additive group of prime order $p$, and let $\mathcal{P}: \{0,1\}^\ell \rightarrow \mathbb{Z}_p$ be a public function. The PCCDH assumption considers a two-stage adversary that operates as: in the first stage, the adversary is given $\left((M, N, X) \in \mathbb{G}^3, \mathcal{P}\right)$ and is required to output a value $Y \in \mathbb{G}$. Then, a random value $k \in \{0,1\}^\ell$ is chosen and the corresponding value $r = \mathcal{P}(k) \in \mathbb{Z}_p$ is computed. In the second stage, the adversary is given $r$ and attempts to compute $K = \text{CDH}(X - r \cdot N,\; Y - r \cdot M)$. The PCCDH assumption states that the advantage of any probabilistic polynomial-time adversary in successfully computing $K$ is negligible. Intuitively, even though the adversary may be able to guess the correct password $k$ (and hence $r$) with some probability in the first stage, it cannot do better than random guessing.
}

\section{Our Proposed Construction}
\label{method}
In this section, we present an overview of our key agreement scheme, specifically designed to address the challenges associated with implementing encryption techniques using physiological signals. Our approach is centered on the use of the unique characteristics of individual physiological signals to generate secure encryption keys.

\subsection{System Overview}
\label{Sec:System Overview}

\begin{figure}[!t]
  \centering
  \includegraphics[width=0.9\linewidth ]{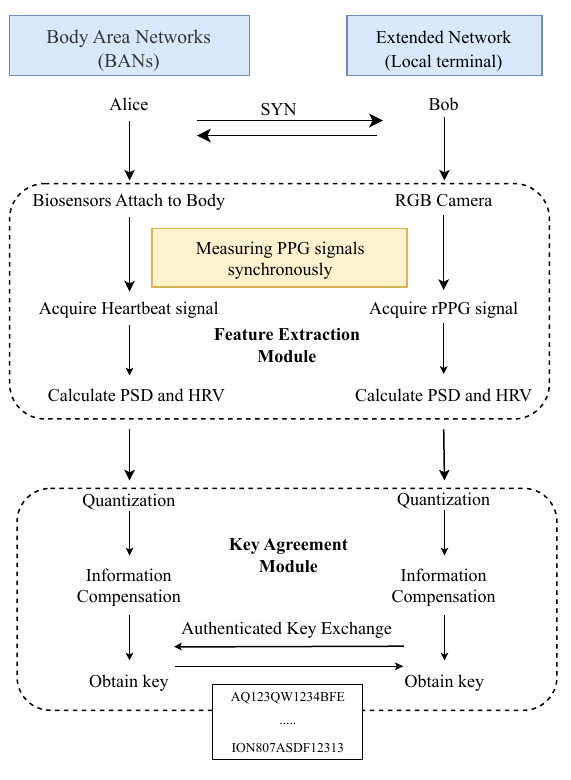}
  \caption{rPPG-based key agreement. Alice stands for BAN. Bob indicates an extended network. It can be any medical device with a camera. Alice and Bob share symmetric keys through the feature extraction and key agreement modules.}
  \label{keyagree}
\end{figure}

Fig.~\ref{keyagree} illustrates our proposed key agreement scheme, where heartbeat-related physiological features are used to facilitate session key generation. In this setting, Alice represents a BAN, while Bob denotes an extended network device, such as a local terminal or a medical monitoring device equipped with a camera. The biosensors in Alice's BAN are assumed to support heartbeat signal acquisition, such as ECG or PPG, whereas Bob is capable of capturing facial videos for rPPG extraction. In addition, both Alice and Bob are assumed to be equipped with wireless communication modules.

The rPPG-based key agreement scheme consists of two main modules: feature extraction and key agreement. In the \textbf{feature extraction module}, the process begins with Alice initiating communication by sending a request message to Bob. Upon receiving Bob's acknowledgment, Alice captures her heartbeat signal in a synchronized manner. Relevant features are then extracted from the acquired signal. Concurrently, Bob calculates comparable features by analyzing the rPPG signals obtained from the faces in the captured videos. Next, in the \textbf{key agreement module}, the extracted physiological features are quantized into binary representations. To mitigate potential discrepancies caused by signal noise or transmission inconsistencies, an error correction mechanism is employed to perform information compensation. Based on the sequences derived from the heartbeat signal on Alice’s side and the rPPG signal on Bob’s side, the two parties can engage in an authenticated key agreement. This process enables Alice and Bob to independently derive a shared secret key.

Given that various regions of the human body can reliably capture similar PPG signals, Alice and Bob can exploit this characteristic to negotiate an identical session key, which can subsequently be used to secure their communications.

\subsection{Feature Extraction Module}

\subsubsection{Signal Acquisition}
Alice employs a biosensor that captures the heartbeat signal. In our scenario, PPG signals are being used as an example. PPG is a non-invasive method that employs light-based sensors, such as photodiodes or cameras, to detect variations in blood volume in the tissue's microvascular bed \cite{xie2025state}. The resulting PPG signals are employed to gauge numerous physiological metrics, such as heart rate, pulse waveform, and blood oxygenation. In addition to traditional oximeters, wearable devices with integrated PPG sensors have become widely available, offering a convenient and accessible option for monitoring PPG signals \cite{Alam10012653}.

Bob needs to extract rPPG signals from the facial video. Similar to PPG signals, rPPG can also measure changes in subcutaneous blood volume \cite{bondarenko2025role}. This measurement is performed by analyzing the subtle color changes (reflectance) or intensity (luminance) of the skin in response to cardiac cycles, captured with a camera or other imaging device.
MediaPipe\footnote{\url{https://google.github.io/mediapipe/}} was used to detect the facial area in the video.

\begin{figure}[!t]
  \centering
  \includegraphics[width=0.9\linewidth ]{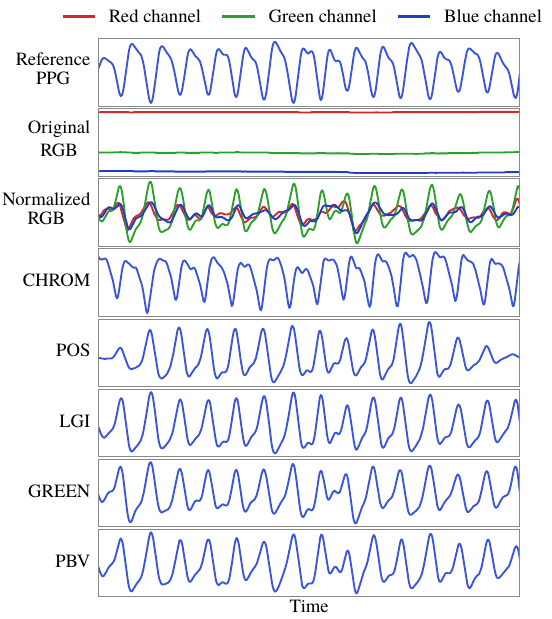}
  \caption{Examples of signals extracted by different rPPG methods. Except for the original RGB signal, all other signals are the result of applying bandpass filtering and detrending.}
  \label{dataexample}
\end{figure}

With MediaPipe, we have acquired the 468 landmarks of the face.
These landmarks help us effectively exclude non-skin regions, such as eyes and eyebrows, which typically contain limited blood information in reflected light.
We derived the original three-channel signals by calculating the average RGB values for all skin regions.
As shown in Fig.~\ref{dataexample}, the original RGB value is calculated from the skin area and exhibits only slight temporal variations.
We then apply an rPPG method to extract the signal. Although several approaches can be used to obtain rPPG signals, we do not adopt deep learning-based rPPG methods in order to limit the computational overhead of the proposed scheme.
To enhance the signal-to-noise ratio, we further employ band-pass filtering to remove frequency components outside the heart rate range.

\begin{figure}[!t]
  \centering
  \includegraphics[width=0.9\linewidth ]{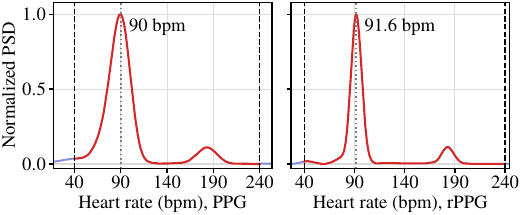}
  \caption{Comparison of power spectral density for PPG and rPPG signals. }
  \label{psdexample}
\end{figure}

\begin{figure}[!t]
  \centering
  \includegraphics[width=0.9\linewidth ]{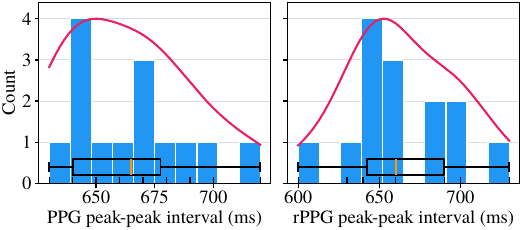}
  \caption{Distribution of the peak-peak intervals of a PPG sample (left) and the corresponding rPPG sample (right).}
  \label{rrexample}
\end{figure}
\subsubsection{Feature Acquisition}
In prior studies, IPI has been utilized to establish a key agreement. However, we found that ambient noise significantly affects the reliability of rPPG signal extracted from the face, resulting in poor correlation with IPI obtained from synchronized PPG signal, typically ranging from 55\% to 70\%.
To overcome this issue, we employ both the frequency and statistical features of the signals. Specifically, we apply a sliding window of length $k$ seconds with a step size of 1~s to extract signal features, and calculate the Power Spectral Density (PSD) of the PPG and rPPG signals, as shown in Fig.~\ref{psdexample}. The frequency component with the highest energy is then selected for key agreement.
For the statistical features, we use the median value of the signal peak-peak interval (PPI). As shown in Fig.~\ref{rrexample}, there is a significant difference between the PPI distributions of the PPG and rPPG signals. Moreover, temporal order is also important when encoding them as feature sequences. Therefore, we use the median value to compensate for such variations. PPI is calculated as the time interval between successive systolic peaks, and the peak detection algorithm in~\cite{elgendi2013systolic} is adopted in this work.

\subsection{Key Agreement Module}

\subsubsection{Feature Synchronisation}

In previous work, signal synchronization between Bob and Alice has been considered essential~\cite{zhang2021h2k}, as inaccurate synchronization may lead to communication failures.
Traditional IPI-based key agreement schemes typically require additional synchronization support, as they rely on the time intervals of successive heartbeat peaks, which cannot be directly synced with timestamps. Therefore, extra synchronization strategies are often needed.
In contrast, our scheme uses a sliding window with a fixed step size, which enables features extracted from both devices to be directly compared according to the same timestamps.

\subsubsection{Feature Quantization}
To facilitate key agreement, the feature values must be transformed into a digital binary sequence suitable for encryption. This conversion is necessary to ensure compatibility with {subsequent cryptographic operations.}
First, we convert the feature values to binary by using Gray code,
{which provides better resilience against small perturbations in the input values. Specifically, the features are quantized and transformed as follows:}

\begin{equation}
g = u \oplus (u \gg 1),
\end{equation}
\begin{equation}
\label{togray}
Bio\_Code = g.zfill(n).
\end{equation}

According to Eq.~(\ref{togray}),
{we apply the standard binary-to-Gray code mapping, where $u$ is the quantized integer feature and $g$ is the resulting Gray-encoded binary value.}
The function $zfill(n)$ fills binary numbers with zeros up to $n$ bits. Based on the possible values of the PSD-based features and PPI-based features, we set the $n$ to 8 and 11, respectively.

This representation ensures that adjacent quantization levels differ by only one bit, making the encoded features robust to small deviations caused by noise or physiological variability. As a result, minor mismatches between the two parties lead only to localized bit errors, which are easier to correct. This property is particularly beneficial for biometric-based key agreement, where the source signals are not strictly identical but remain highly correlated.
After encoding, Alice and Bob rectify the mismatched biometric codes using error correction codes. For example, Turbo codes \cite{berrou1996near} can attain an error rate of $10^{-6}$ or less, signifying that a mere one bit out of every million bits transmitted is prone to be received in error.

\begin{figure}[t]
\centering
{
\setlength{\fboxsep}{0pt}
\setlength{\fboxrule}{0.45pt}

\resizebox{0.5\textwidth}{!}{%
\fbox{%
\procedure{
\vspace{0.1cm}
~~Public parameters: $ID_a, ID_b, \mathbb{G}, P\in\mathbb{G}, H, H_1, H_2$; ~Biometric code: $\widetilde{bc}$.}
{
\quad \quad \textbf{Alice} \> \> \quad \textbf{Bob} \\
\quad a\overset{R}{\leftarrow}\mathbb{Z}_p; \> \> b\overset{R}{\leftarrow}\mathbb{Z}_p; \\
\quad M \leftarrow H(ID_b)\in \mathbb{G}; \> \> N \leftarrow H(ID_a)\in \mathbb{G};\\
\quad X \leftarrow a\cdot P + \widetilde{bc}\cdot M. \> \> Y \leftarrow b\cdot P + \widetilde{bc}\cdot N. \\
\> \sendmessageright*[1.5cm]{ID_a, X} \> \\
\> \sendmessageleft*[1.5cm]{ID_b, Y} \> \\
\quad N \leftarrow H(ID_a); \> \> M \leftarrow H(ID_b);\\
\quad B \leftarrow {Y}-{\widetilde{bc}\cdot N}; \> \> A \leftarrow {X}-{\widetilde{bc}\cdot M};\\
\quad K \leftarrow a\cdot B; \> \> K \leftarrow b\cdot A;\\
\quad sk_a \leftarrow H_1(ID_a,ID_b,A,B, K); \> \> sk_b \leftarrow H_1(ID_a,ID_b,A,B, K); ~~\\
\quad {\mathsf{Auth_a} \leftarrow H_2(ID_a,ID_b,sk_a,1).} \> \> {\mathsf{Auth_b} \leftarrow H_2(ID_a,ID_b,sk_b,2).}~~\\
\> \sendmessageright*[1.5cm]{{\mathsf{Auth_a}}} \> \\
\> \sendmessageleft*[1.5cm]{{\mathsf{Auth_b}}} \> \\
\quad {\mathsf{Auth_b}' \leftarrow H_2(ID_a,ID_b,sk_a,2);} \> \> {\mathsf{Auth_a}' \leftarrow H_2(ID_a,ID_b,sk_b,1);}~~\\
\quad {\text{if}~\mathsf{Auth_b} = \mathsf{Auth_b}',} \> \> {\text{if}~\mathsf{Auth_a} = \mathsf{Auth_a}',}\\
\quad {ssk \leftarrow sk_a.} \> \> {ssk \leftarrow sk_b.} \pclb
}%
}
}
}
\caption{Authenticated Key Agreement (AKA). {It works in a group $\mathbb{Z}_p$, and an additive group $\mathbb{G}$ with order $p$ where the discrete log problem is hard. The identifiers $ID_a$ and $ID_b$, generator $P$, hash functions $H$, $H_1$, $H_2$ are exchanged in the initiate communication as mentioned in Section~\ref{Sec:System Overview}.  The biometric code $\widetilde{bc}$ is the synchronized biometric code, which is derived from aligned physiological signals.} }
\label{fig:ake}
\end{figure}

\subsubsection{Authenticated Key Agreement}

The synchronized biometric code (represented by $\widetilde{bc}$) can then be utilized to authenticate each other and negotiate a session key.  As shown in Fig.~\ref{fig:ake}, Alice and Bob select $a\overset{R}{\leftarrow}\mathbb{Z}_p$ and $b\overset{R}{\leftarrow} \mathbb{Z}_p$, respectively. Then, Alice computes $X=a\cdot P+ \widetilde{bc}\cdot M$ and sends the $X$ to Bob. At the same time, Bob computes $Y=b\cdot P+ \widetilde{bc}\cdot N$ and sends the $Y$ to Alice. Next, Alice can recover a parameter $B={b}\cdot P=Y-\widetilde{bc}\cdot N$ while Bob can recover a parameter $A={a}\cdot P=X-\widetilde{bc}\cdot M$. Finally, Alice and Bob can compute a session key by calculating $sk_a=H_1(ID_a,ID_b,A,B,K)$ and $sk_b=H_1(ID_a,ID_b,A,B,K)$, respectively. {To further ensure the consistency of the derived session key, we follow the generic construction of authenticated key exchange~\cite{vos2025hybrid} and add a lightweight mutual authentication step. Specifically, the session key is accepted only if the received authentication tag equals the locally recomputed one, which implies $sk_a = sk_b$. This verification guarantees that both parties have derived the same key from consistent biometric inputs and ephemeral values. Note that in each session, both Alice and Bob independently select fresh ephemeral values $a, b \overset{R}{\leftarrow} \mathbb{Z}_p$. As a result, the derived session key $ssk$ is guaranteed to be unique for each session.}

\section{Security Analysis}

\label{securitysec}

In this section, we analyze the security of SigK. We first provide a formal proof in the random oracle model, showing that the proposed scheme achieves AKA security under the CDH and PCCDH assumptions. We then provide informal security analysis to further demonstrate that the proposed SigK satisfies common security properties.

\subsection{Security Model}

Our security proof is based on the BPR~\cite{vos2025hybrid} model. The security model is defined as follows.

\textit{Participant}:  The system involves three entities, namely Alice $(A)$, Bob $(B)$, and a probabilistic polynomial-time adversary $\mathcal{A}$. Alice represents a body-area network device that can acquire heartbeat-related signals, such as ECG or PPG. Bob represents an external device that can capture facial video and extract rPPG signals. Each participant may have multiple protocol instances. The $i$-th instance of Alice and the $j$-th instance of Bob are denoted by $A^i$ and $B^j$, respectively. The adversary $\mathcal{A}$ is assumed to fully control the public communication channel between Alice and Bob, and can eavesdrop, modify, replay, delay, or inject messages. However, $\mathcal{A}$ cannot directly access the private physiological measurements obtained by Alice's sensor or Bob's local rPPG extraction process. Generally, $\mathcal{A}$ is able to control communication channels between $A$ and $B$ and make queries to the following oracles.

\begin{itemize}[leftmargin=*]
\item[-] \textit{Execute}: The passive attacks are captured by this query, and its output comprises all communication transcripts of the protocol between $A$ and $B$.

\item[-] \textit{Send}: The active attacks are captured by this query. $\mathcal{A}$ forges a message $m$, and then sends $m$ to instance $I^i$ ($I\in\{A,B\}$) and obtains the response from $I^i$.

\item[-] \textit{Reveal}: The misuse of session keys is modelled by this query. It responds with
 the session key $sk$ of instance $I^{i}$ to $\mathcal{A}$ if the instance $I^i$ accepts the session. Otherwise, it returns $\perp$.

\item[-] \textit{Test}: Upon receiving the query, a coin $\beta$ is flipped. If $\beta=0$, a random value with the same size as $sk$ is returned to $\mathcal{A}$. If $\beta=1$, then the real $sk$ is returned to $\mathcal{A}$. This query can only be called once for a fresh instance.
\end{itemize}

\textit{Freshness.} An accepted instance is fresh if neither the instance nor its matching session has been queried to \textit{Reveal}. Freshness excludes trivial cases in which the adversary directly obtains the target session key. In the biometric-assisted setting, a fresh session also requires that the adversary does not obtain the private physiological feature sequences used by both parties to derive the target session key.

\textit{AKA-Security}. Let $\mathsf{Succ}(\mathcal{A})$ refer to the event that $\mathcal{A}$ asks a $\textit{Test}\left(I^i\right)$ query for a fresh instance and outputs $\beta$ correctly. The advantage of an adversary for breaching the semantic security of an authenticated key agreement scheme is presented as $\mathsf{Adv}^{AKA}(\mathcal{A})=|2 \operatorname{Pr}[\mathsf{Succ}(\mathcal{A})]-1|$. SigK achieves the AKA-security if for any PPT adversary $\mathcal{A}$, the advantage $\mathsf{Adv}^{AKA}(\mathcal{A})$ is negligible.

\subsection{ Security Proof}

Here, we prove that if the CDH assumption and the PCCDH assumption hold in $\mathbb{G}$ and the biometric code has sufficient entropy, then an adversary $\mathcal{A}$ that issues at most $q_H$ random-oracle queries and $q_s$ \textit{Send} queries has only a negligible advantage $\mathsf{Adv}^{\mathsf{AKA}}(\mathcal{A})$ to break the AKA security of SigK.

Our proof consists of a sequence of hybrid games, starting with the real attack and ending in a game where the adversary’s advantage is negligible.

\textit{Game} $E_0$: This game simulates the real attack. Let $\mathsf{Succ}_i$ be the event that the attacker successfully guesses the bit $\beta$ in the \textit{Test} query in Game $E_i$, then we have $\mathsf{Adv}^{AKA}(\mathcal{A})=\left|2\operatorname{Pr}\left[\mathsf{Succ}_0\right]-1\right|.$

\begin{figure}[t]
\centering
\small

\setlength{\fboxsep}{6pt}
\setlength{\fboxrule}{0.45pt}

\newcommand{\oracleline}{%
  \vspace{-3pt}
  \par\noindent
  \hspace*{-\fboxsep}%
  \rule{\dimexpr\linewidth+2\fboxsep\relax}{0.45pt}%
  \par
  \vspace{1pt}
}

\fbox{%
\parbox{0.93\linewidth}{%
\setlength{\parskip}{0pt}
\setlength{\parsep}{0pt}

\begin{itemize}[leftmargin=*,itemsep=0pt,topsep=0pt,parsep=0pt,partopsep=0pt]
\item[-] On a query $H_1(q)$, if there exists a record $(q,k)$ in $\mathcal{L}_h$, return $k$; otherwise, randomly choose $k \leftarrow \{0,1\}^{\ell}$, record $(q,k)$, and return $k$.
\end{itemize}

\oracleline

- On a query \textit{Execute} ($A^i, B^{j}$), processed as follows:
\begin{itemize}[leftmargin=*,itemsep=0pt,topsep=0pt,parsep=0pt,partopsep=0pt]
    \item[] $a \gets \mathbb{Z}_p, M \gets H(ID_b), X \leftarrow a\cdot P + \widetilde{bc}\cdot M;$
    \item[] $b \gets \mathbb{Z}_p, N \gets H(ID_a), Y \leftarrow b\cdot P + \widetilde{bc}\cdot N;$
    \item[] $K=a\cdot b\cdot P;$
    \item[] $sk\gets H_1(ID_a,ID_b,a\cdot P,b\cdot P,K)$;
    \item[] $\mathsf{Auth_a} \gets H_2(ID_a,ID_b, sk, 1)$;
    \item[] $\mathsf{Auth_b} \gets H_2(ID_a,ID_b, sk, 2)$;
    \item[] $ssk \gets sk;$
    \item[] {return} $\left((ID_a, X), (ID_b, Y), \mathsf{Auth_a}, \mathsf{Auth_b}\right)$.
\end{itemize}

\oracleline
- On a query \textit{Send} ($A^{i}$, start), proceed as follows:
\begin{itemize}[leftmargin=*,itemsep=0pt,topsep=0pt,parsep=0pt,partopsep=0pt]
 \item[] if ActiveSessionIndex$\neq 0$, then abort;
 \item[] else ActiveSessionIndex$=i$;
 \item[] $a \gets \mathbb{Z}_p, M \gets H(ID_b), X \leftarrow a\cdot P + \widetilde{bc}\cdot M;$
 \item[] {return} $(ID_a, X)$.
\end{itemize}

- On a query \textit{Send} $\left(B^{i}, (ID_a, X)\right)$, proceed as follows:
\begin{itemize}[leftmargin=*,itemsep=0pt,topsep=0pt,parsep=0pt,partopsep=0pt]
  \item[] $b \gets \mathbb{Z}_p, N \gets H(ID_a), Y \leftarrow b\cdot P + \widetilde{bc}\cdot N;$
  \item[] $M \gets H(ID_b)$;
  \item[] $K\gets b\cdot\left({X}-{\widetilde{bc}\cdot M}\right)$;
  \item[] $sk_b \gets H_1\left(ID_a,ID_b, \left({X}-{\widetilde{bc}\cdot M}\right), b\cdot P, K\right);$
  \item[] {return} $(ID_b, Y)$.
\end{itemize}

- On a query \textit{Send} $\left(A^{i}, (ID_b, Y)\right)$, proceed as follows:
\begin{itemize}[leftmargin=*,itemsep=0pt,topsep=0pt,parsep=0pt,partopsep=0pt]
  \item[] $N \gets H(ID_a)$;
  \item[] $K\gets a\cdot\left({Y}-{\widetilde{bc}\cdot N}\right)$;
  \item[] $sk_a \gets H_1\left(ID_a,ID_b,a\cdot P,\left({Y}-{\widetilde{bc}\cdot N}\right), K\right);$
\end{itemize}

- On a query \textit{Send} $\left(A^{i}, \mathsf{Auth_b}\right)$, proceed as follows:
\begin{itemize}[leftmargin=*,itemsep=0pt,topsep=0pt,parsep=0pt,partopsep=0pt]
  \item[] $\mathsf{Auth_b}'\gets H_2\left(ID_a,ID_b, sk_a, 2\right)$;
  \item[] {if} $\mathsf{Auth_b}'$==$\mathsf{Auth_b}$, {then} $ssk \gets sk_a$;
  \item[] ActiveSessionIndex=0.
\end{itemize}

- On a query \textit{Send} $\left(B^{i}, \mathsf{Auth_a}\right)$, proceed as follows:
\begin{itemize}[leftmargin=*,itemsep=0pt,topsep=0pt,parsep=0pt,partopsep=0pt]
  \item[] $\mathsf{Auth_a}'\gets H_2\left(ID_a,ID_b, sk_b, 1\right)$;
  \item[] {if} $\mathsf{Auth_a}'$==$\mathsf{Auth_a}$, {then} $ssk \gets sk_b$;
  \item[] ActiveSessionIndex=0.
\end{itemize}

\oracleline
- On a query \textit{Reveal} ($I^i$), proceed as follows:
\begin{itemize}[leftmargin=*,itemsep=0pt,topsep=0pt,parsep=0pt,partopsep=0pt]
 \item[] {if} $ssk$ is defined for instance $I^i$
 \item[] {then return} $ssk$;
 \item[] {else return} $\perp$.
\end{itemize}

\oracleline
- On a query \textit{Test} ($I^i$), proceed as follows:
\begin{itemize}[leftmargin=*,itemsep=0pt,topsep=0pt,parsep=0pt,partopsep=0pt]
 \item[] $\beta \gets \{0,1\};$
 \item[] {if} $\beta=0$, randomly choose a $r\in\{0,1\}^{\ell}$, {return} $r$;
 \item[] {else return} $ssk\gets$\textit{Reveal} ($I^i$).
\end{itemize}
}%
}
\caption{Simulation of $H_1$, \textit{Execute}, \textit{Send}, \textit{Reveal}, and \textit{Test} oracles.}
\label{fig:sim}
\end{figure}

\textit{Game} $E_1$: This game simulates the $H_1$, \textit{Execute}, \textit{Reveal}, and \textit{Send} oracles following Fig.~\ref{fig:sim}. Since all oracles are simulated as the real attack, this game cannot be distinguished from Game $E_0$. Thus, we have $\Pr\left[\mathsf{Succ}_1\right]=\Pr\left[\mathsf{Succ}_0\right]$.

\textit{Game} $E_2$: This game is identical to Game $E_1$, except that the simulation is aborted if a collision occurs among random-oracle outputs or if the adversary successfully guesses the synchronized biometric code used in a fresh session. By the birthday bound, the probability of a random-oracle output collision is at most $q_H^2/2^{\ell}$. Moreover, assuming that the synchronized biometric code $\widetilde{bc}$ has conditional min-entropy at least $\kappa$ given all public information available to the adversary, the probability of guessing $\widetilde{bc}$ is at most $2^{-\kappa}$ for each session. Therefore, over at most $q_s$ sessions, this probability is bounded by $q_s/2^{\kappa}$. Thus, $|\Pr(\mathsf{Succ}_2)-\Pr(\mathsf{Succ}_1)|\leq q_H^2/2^{\ell}+q_s/2^{\kappa}$, which is negligible for sufficiently large $\ell$ and $\kappa$.

\textit{Game} $E_3$: This game is identical to Game $E_2$, except that the session key of each fresh accepted session is replaced by an independent random value. For a matching session between $A^i$ and $B^j$, the simulator defines the session transcript as $sid=(ID_{a},ID_{b},X,Y)$. In the real execution, both parties recover $A=X-\widetilde{bc}\cdot M=aP$ and $B=Y-\widetilde{bc}\cdot N=bP$, compute the Diffie--Hellman component $K=abP$, and derive the session key as $sk=H_1(ID_{a},ID_{b},A,B,K)$. In Game $E_3$, the simulator instead samples an independent value $\widehat{sk}\leftarrow\{0,1\}^{\ell}$ for this session and uses $\widehat{sk}$ in all subsequent computations of authentication tags and the final session key.

Let $\mathsf{AskH}_1^{\mathsf{match}}$ denote the event that the adversary queries $H_1$ on the exact tuple $(ID_{a},ID_{b},A,B,K)$ associated with a fresh matching session, where $A=aP$, $B=bP$, and $K=abP$. If $\mathsf{AskH}_1^{\mathsf{match}}$ does not occur, then the value $H_1(ID_{a},ID_{b},A,B,K)$ is sampled by the random oracle as a uniformly random value and remains independent of the adversary's view. Therefore, replacing it with the independently sampled value $\widehat{sk}$ does not change the adversary's view.

Hence, Games $E_2$ and $E_3$ can be distinguished only if $\mathsf{AskH}_1^{\mathsf{match}}$ occurs. To trigger $\mathsf{AskH}_1^{\mathsf{match}}$, the adversary must query $H_1$ with the correct Diffie--Hellman component $K=abP$ corresponding to $A=aP$ and $B=bP$. If the adversary can issue such a query with non-negligible probability, then one can construct a CDH solver that embeds a CDH challenge $(P,U,V)$ into one guessed matching session by setting the concealed messages as $X=U+\widetilde{bc}\cdot M$ and $Y=V+\widetilde{bc}\cdot N$. Once the adversary queries $H_1$ with the corresponding tuple $(ID_{a},ID_{b},U,V,K)$, the solver outputs $K$ as the solution to the CDH challenge. Since the adversary can make at most $q_s$ \textit{Send} queries, we have $|\Pr(\mathsf{Succ}_3)-\Pr(\mathsf{Succ}_2)|\leq q_s\cdot \mathsf{Adv}^{\mathsf{CDH}}_{\mathbb{G}}(\lambda)$.

\textit{Game} $E_4$: This game handles active attacks modeled by the $\mathsf{Send}$ oracle. In such attacks, the adversary may inject, modify, replay, or replace protocol messages. Compared with Game $E_3$, this game additionally aborts if a fresh honest instance accepts a session for which no matching session exists. Let $\mathsf{Forge}$ denote this event.

We next bound the probability of $\mathsf{Forge}$. For a non-matching accepted session, the accepting instance computes its session key from a concealed Diffie--Hellman component. Let $r=\widetilde{bc}$. According to the proposed scheme, the protocol messages are of the form $X=aP+rM$ and $Y=bP+rN$. Since the honest instance knows $r$, it recovers $A^\ast=X-rM$ and $B^\ast=Y-rN$, and then computes the hidden Diffie--Hellman component $K^\ast=\mathsf{CDH}(A^\ast,B^\ast)$. The session key is derived as $sk=H_1(ID_{a},ID_{b},A^\ast,B^\ast,K^\ast)$.

Let $\mathsf{AskH}_1^{\mathsf{act}}$ denote the event that the adversary queries $H_1$ on the exact tuple $(ID_{a},ID_{b},A^\ast,B^\ast,K^\ast)$ corresponding to such a non-matching active session. If this event occurs with non-negligible probability, then one can construct a solver for the PCCDH problem. Specifically, the solver embeds the PCCDH challenge into the concealed protocol message of an honest instance and uses the adversary's $\mathsf{Send}$ query as the chosen active message. Once the adversary queries $H_1$ with the correct tuple containing $K^\ast=\mathsf{CDH}(A^\ast,B^\ast)$, the solver extracts $K^\ast$ from this query and outputs it as the solution to the PCCDH challenge. Therefore, up to the guessing loss for the target active session, we have $\Pr(\mathsf{AskH}_1^{\mathsf{act}})\leq q_s\cdot \mathsf{Adv}^{\mathsf{PCCDH}}_{\mathbb{G}}(\lambda)$, where $q_s$ denotes the number of active sessions generated through $\mathsf{Send}$ queries.

If $\mathsf{AskH}_1^{\mathsf{act}}$ does not occur, then the session key $sk$ of the non-matching active session remains hidden from the adversary. Therefore, to make an honest instance accept, the adversary must either query $H_2$ with the correct hidden session key or directly guess a valid authentication tag. Since the session key has length $\ell$ and the authentication tag has length $\ell_{auth}$, the probability of such a forgery is at most $q_{H_2}/2^{\ell}+q_s/2^{\ell_{auth}}$.
Consequently, $|\Pr(\mathsf{Succ}_4)-\Pr(\mathsf{Succ}_3)|\leq q_s\cdot \mathsf{Adv}^{\mathsf{PCCDH}}_{\mathbb{G}}(\lambda)+q_{H_2}/ 2^{\ell}+q_s/ 2^{\ell_{auth}}$.

\vspace{0.1cm}
Combining the above games, we can observe that the advantage $\mathsf{Adv}^{AKA}(\mathcal{A})$ is negligible as the CDH and PCCDH assumptions hold in $\mathbb{G}$ and the collision probabilities of the random oracles and the biometric code
are negligible. Therefore, SigK achieves AKA security.

\subsection{{Other Security Analysis}}
Our scheme also meets other security properties which are discussed informally.

{
\textit{Confidentiality:}
The proposed scheme ensures the confidentiality of communication by establishing a session-specific symmetric key that is known only to the two authenticated parties. All subsequent messages exchanged between Alice and Bob are encrypted using this session key. Since the key is derived from ephemeral randomness and biometric input, an adversary who does not possess the session key cannot decrypt or infer the content of the communication.}

\textit{Authenticity:}
The exchange and verification of $\mathsf{Auth}_a$ and $\mathsf{Auth}_b$ ensures that both parties have derived the same session key from matching biometric features and ephemeral values. A successful verification confirms that the communicating peer is legitimate and possesses consistent keying material. Conversely, any discrepancy in the biometric code $\widetilde{bc}$ or the ephemeral randomness results in authentication failure, thereby preventing impersonation and session hijacking.

{
\textit{Man-in-the-Middle (MitM) Attack Resistance:}
The proposed scheme resists MitM attacks by performing a mutual authentication step after the session key generation. Specifically, the final authentication factors $\mathsf{Auth}_a$ and $\mathsf{Auth}_b$ are calculated on the agreed key, ensuring that only parties with consistent views of the session can complete the process. An adversary who intercepts messages ($X'$, $Y'$, $\mathsf{Auth}_a'$, or $\mathsf{Auth}_b'$) cannot succeed, as the adversary lacks knowledge of the session-specific biometric code $\widetilde{bc}$, and thus cannot compute a valid session key. Consequently, any message modification or injection by a third party will lead to the failure of key agreement and the termination of the current session. }

{
\textit{Forward Secrecy:}
The session key in the proposed scheme is derived from fresh ephemeral values ($a$,$b$), which are randomly generated for each session. This guarantees that every session results in a unique and independent session key. Importantly, the key derivation process does not rely solely on static identifiers or long-term secrets, but rather incorporates session-specific randomness. Therefore, even if the biometric code $\widetilde{bc}$ or the current session key is compromised, an adversary cannot retroactively compute or recover previously established session keys, thereby providing forward secrecy.}

{
\textit{Precomputation Attack Resistance:}
A precomputation attack refers to an adversary who attempts to compromise the scheme by generating a lookup table of the most probable secret keys. However, in our scheme, the biometric signals contribute session-specific entropy that is inherently unpredictable, and the $\widetilde{bc}$ is time-varying and highly sensitive to measurement noise, which prevents the adversary from constructing a meaningful dictionary or reuse of previously observed data. As a result, each session remains resilient against offline guessing and precomputation-based attacks.}

\section{Evaluations }
\label{evaluations}
In this section, we evaluate the proposed rPPG-based key agreement scheme. We conducted a series of experiments on five common rPPG methods (CHROM~\cite{de2013robust}, POS~\cite{wang2016algorithmic}, LGI~\cite{Pilz8575328}, GREEN~\cite{verkruysse2008remote}, PBV~\cite{de2014improved}) on three datasets (PURE~\cite{stricker2014non}, UBFC\_rPPG~\cite{bobbia2019unsupervised}, LGI\_PPGI~\cite{Pilz8575328}). We use PyVHR\footnote{\url{https://github.com/phuselab/pyVHR}} \cite{boccignone2022pyvhr} to implement various rPPG methods to ensure uniform analysis.

\begin{figure*}[!t]
  \centering
  \includegraphics[width=0.9\linewidth ]{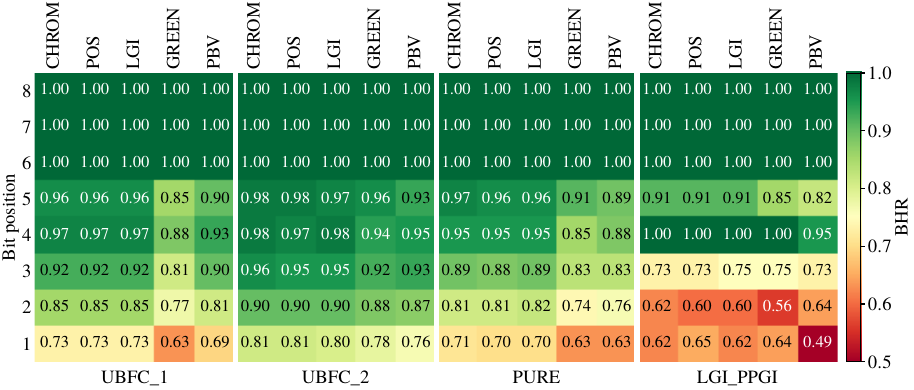}
  \caption{PSD-based bit hit rate for different bit positions on the four evaluation subsets.  The size of the sliding window is set to 12~s. A higher BHR value indicates a more accurate system.}
  \label{psdbithit}
\end{figure*}

\begin{table}[ht]
\centering
\caption{\centering Overview of Datasets Used}
\renewcommand\arraystretch{1.2}
\setlength{\tabcolsep}{2.7mm}{
\begin{tabular}{l|l|l|l}
\toprule
\textbf{Dataset} & \textbf{Subjects}  & \textbf{FPS} & \textbf{Resolution} \\ \hline
PURE             & 10                    & 30                        & $640\times480$      \\ \hline
UBFC\_rPPG       & 42                  & 30                        & $640\times480$      \\ \hline
LGI\_PPGI        & 25                & 25               & $640\times480$      \\ \hline
\end{tabular}}
\label{tab:dataset_info}
\end{table}

\subsection{Datasets}
\begin{itemize}
  \item  \textbf{PURE}: This dataset contains a collection of 10 individuals performing six different controlled head movements in front of a camera. Each video lasts one minute with a frame rate of 30 FPS and a resolution of $640\times480$. Parallel to the video recordings, reference PPG signals were obtained using a finger clip pulse oximeter with a sampling rate of 60~Hz.

  \item \textbf{UBFC\_rPPG}: It consists of a video in uncompressed 8-bit RGB format running at 30 FPS with a resolution of $640\times480$, and the corresponding PPG recording. The database is comprised of two sub-datasets. The first sub-dataset (UBFC\_1) consists of 8 videos, with the subjects being asked to sit still. The second sub-dataset (UBFC\_2) involves the subjects playing a time-sensitive mathematical game designed to increase their heart rate. Records of 42 participants are shared.

  \item \textbf{LGI\_PPGI}: It captured video and PPG signals from 25 participants. The data was collected over four different sessions (resting, talking, rotation, and gym), each designed to capture different types of facial and head movements and illumination conditions. The camera video stream was captured in uncompressed format with an average frame rate of 25 FPS, $640\times480$ resolution.

\end{itemize}

\subsection{Evaluation Metrics}

In our work, the Bit Hit Rate (BHR) and the Key Generation Rate (KGR)~\cite{zhang2021h2k} are adopted as the primary evaluation metrics, since the main objective of this study is to examine the feasibility of rPPG-based key agreement. Specifically, BHR measures the bit-level consistency between the key materials generated from contact-based PPG and contactless rPPG signals, whereas KGR quantifies the number of key bits that can be derived per unit time from physiological observations. Consequently, these two metrics provide direct evidence for evaluating whether contactless rPPG signals can support reliable and practical key agreement in mHealth scenarios.

Formally, BHR is defined as the proportion of identical bits between two biometric codes generated from the same physiological source. A higher BHR indicates stronger consistency between the key materials derived by the two parties, which is essential for successful key agreement. Let $\odot$ denote the bit-wise XNOR operation, the BHR is calculated as follows:
\begin{equation*}
tmp = Bio\_Code_{PPG} \odot  Bio\_Code_{rPPG},
\end{equation*}
\begin{equation}
BHR = \frac{Number\ of\ 1\ in\ tmp}{len(tmp)}.
\end{equation}

KGR measures the number of key bits that can be generated per unit time, and is used to evaluate the efficiency of the key generation process. In our setting, KGR reflects the rate at which usable key materials can be derived from physiological observations. It is defined as follows:
\begin{equation}
KGR = \frac{len(Bio\_Code_{rPPG})}{T},
\end{equation}
where $len(\cdot)$ denotes the number of bits and $T$ denotes the time in seconds to generate the $Bio\_Code_{rPPG}$.

\begin{figure*}[!t]
  \centering
  \includegraphics[width=0.9\linewidth ]{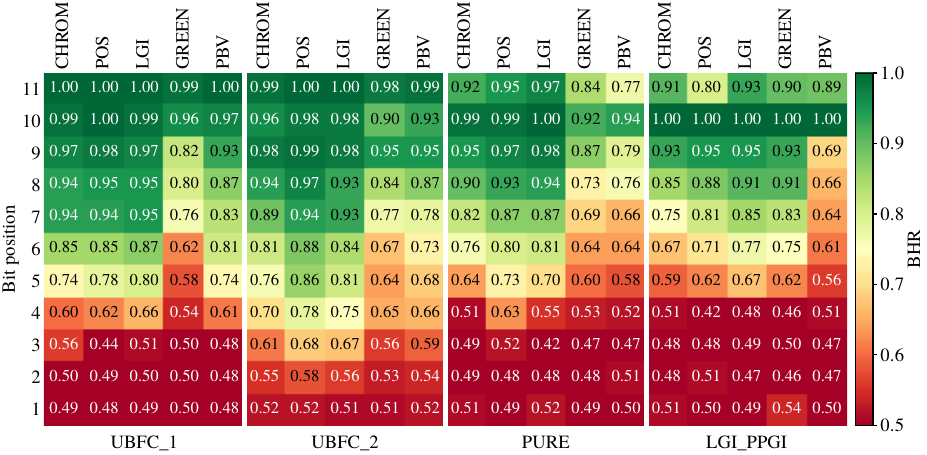}
  \caption{PPI-based bit hit rate for different bit positions on the four evaluation subsets. The size of the sliding window is set to 12~s.}
  \label{bitpositionhrv}
\end{figure*}

\subsection{Selection of Bit Positions}

Finding the appropriate bit positions is essential in the biometric key exchange process as it defines the key value we generate from the biometric features. Selecting the most significant bits (MSBs) for key generation resulted in a robust key but compromised the Bit Hit Rate (BHR).
Conversely, using less significant bits (LSBs) improved the BHR but made the key more predictable and less secure, highlighting a critical trade-off in biometric systems.
A high BHR for the selected bit positions is desirable as it indicates a more accurate biometric match between devices. Our empirical analysis, illustrated in Fig.~\ref{psdbithit} and Fig.~\ref{bitpositionhrv}, demonstrates a clear trend: as the bit position decreases, so does the BHR.
This trend suggests that lower bit positions, while contributing to a higher BHR, may not offer the necessary robustness for secure key generation.

Furthermore, ensuring that the values of the selected bits occur as randomly as possible is crucial for maintaining the security and robustness of the key generation systems. In IPI-based key generation, for instance, the location of the selected bits determines the randomness of the extracted features \cite{Rostami2516658}.
Our goal is to achieve a distribution of the extracted bits' values that is as close to random as possible, meaning that the feature values are evenly distributed under specific bit positions. This randomness is key to thwarting predictability in the generated biometric keys.

\begin{figure*}[!t]
  \centering
  \includegraphics[width=0.9\linewidth ]{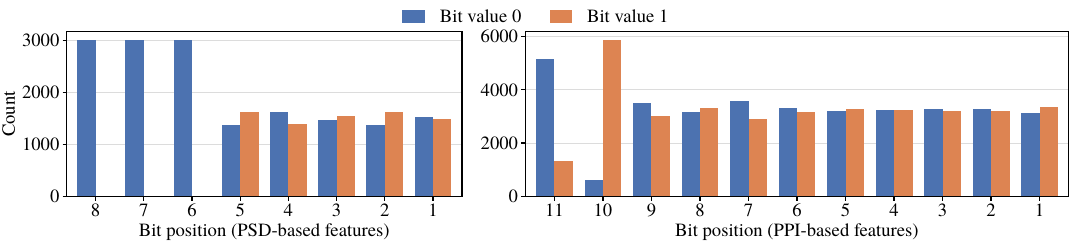}
  \caption{Distribution of the bit values of the PSD-based (left) and PPI-based (right) features at different bit positions over the four evaluation subsets. We discarded bits 8, 7, and 6 of the PSD-based features and bits 11 and 10 of the PPI-based features due to uneven distribution.}
  \label{bitdistr}
\end{figure*}

To ensure this randomness, we meticulously checked the distribution of the values of each bit position. However, our findings, as presented in Fig.~\ref{bitdistr}, revealed significant inconsistencies. For the PSD-based features, Bit-8, 7, and 6 consistently showed a value of 0, indicating no variability.
Similarly, for the PPI-based features, Bit-11 and Bit-10 exhibited a predominance of the values 0 and 1, respectively, with more than 80\% occurrence. These positions often resulted in similar patterns across different individuals, thereby increasing the predictability of the biometric key and exposing a potential security flaw.
Such grossly uneven distributions not only compromise the uniqueness of the biometric keys but also imply that an attacker can easily predict the bit value of these positions using basic statistical methods.
Given these vulnerabilities, and to enhance the security and reliability of our biometric key exchange system, we have decided to exclude the bit values in these positions from our subsequent experiments.
After excluding these bit values, it is not difficult to observe that the PPI-based method has a significant advantage over PSD in terms of KGR (as shown in Fig.~\ref{bitdistr}). This observation suggests that PPI-based approaches are more effective in generating secure and reliable biometric keys under the constraints of our modified criteria.
Additionally, we observed that the performance of various methods in the LGI\_PPGI dataset was generally lower compared to other datasets. We attribute this primarily to the increased amount of head movement among participants in the LGI\_PPGI dataset. Head movements introduce greater complexity to the data.  The movement of the head alters the facial lighting, introducing more noise into the data.

\begin{figure*}[!t]
  \centering
  \includegraphics[width=0.9\linewidth ]{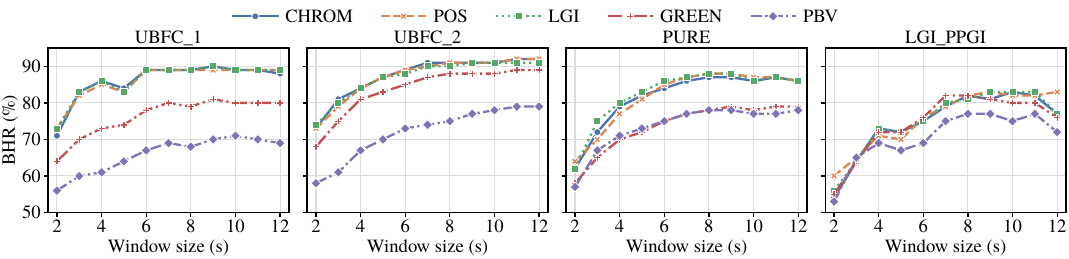}
  \caption{PSD-based bit hit rate for different sliding window sizes {on the four evaluation subsets}. The BHR increases as the window size increases until it plateaus after the window size reaches 7.}
  \label{winsize}
\end{figure*}

\subsection{Performance of different window sizes}

\begin{figure*}[!t]
  \centering
  \includegraphics[width=0.9\linewidth ]{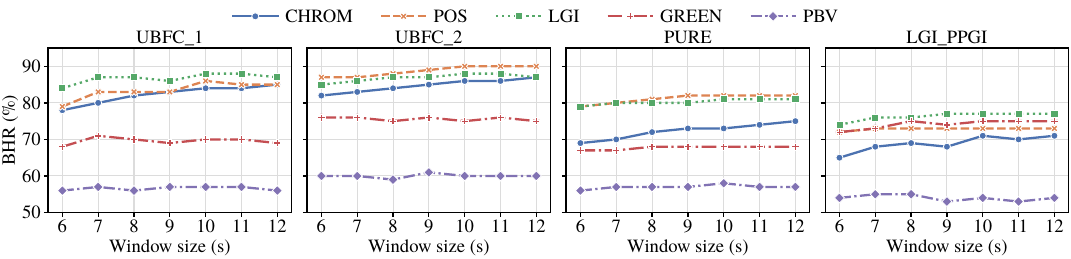}
  \caption{PPI-based bit hit rate for different sliding window sizes {on the four evaluation subsets}. The BHR does not change dramatically with the increase in window size.}
  \label{winsizehrv}
\end{figure*}

To determine the optimal size for the sliding window, we comprehensively evaluated the BHR of various schemes utilizing different window sizes.
For the performance of PSD-based features, we explored window sizes ranging from 2 to 12. As depicted in Fig.~\ref{winsize}, a notable trend emerges: the BHR gradually increases with the window size. However, upon reaching a window size of 7, the growth in BHR plateaus, suggesting an optimal balance between window size and performance efficiency. This phenomenon could be attributed to larger windows providing more data for analysis, thereby enhancing the accuracy of PSD-based feature extraction. However, beyond a certain size, the incremental gains in BHR diminish, possibly due to the inclusion of redundant or less relevant information.

Among the five rPPG methods we analyzed, the POS method consistently demonstrated the most stable performance across various datasets, underscoring its robustness. While CHROM and LGI methods also achieved high BHR, their performance notably declined in the LGI\_PPGI dataset when the window size was increased to 12. This decline could be indicative of the methods' sensitivity to specific dataset characteristics, such as motion artifacts or lighting variations that are more pronounced in larger window sizes.

For performance based on PPI-based features, we explored window sizes from 6 to 12. The starting point of 6 was chosen based on the requirement to acquire sufficient PPI information, which entails at least two PPI values over three heartbeat cycles. Contrary to the PSD-based features, the BHR for PPI-based methods, as shown in Fig.~\ref{winsizehrv}, does not exhibit an increase with larger window sizes. This finding suggests that PPI features might be more dependent on the quality rather than the quantity of data within a window. The POS and LGI methods again showed leading BHRs across datasets in the PPI-based evaluation. This consistent performance points to their potential for biometric key generation, especially in applications where the window size is constrained or where rapid processing is a priority.

\section{Discussion}
\label{discussion}
In our method, some generated bits are completely random, and some are fixed. After discarding invalid bits that do not contribute to security, we find that the PSD-based method can extract five valid bits within seven heartbeat cycles. Similarly, the PPI-based method allows extracting five valid bits in just six heartbeat cycles. Considering an adult's resting heart rate ranges between 60 and 100 beats per minute \cite{kang2017association}, our scheme facilitates the generation of a 256-bit key within an acceptable 5--6 minutes in a medical setting. {This latency is consistent with typical signal acquisition durations during patient monitoring and does not interfere with standard clinical workflows.}
{Once the initial key is established, the dynamic nature of the underlying physiological signals enables periodic regeneration of fresh keys using newly collected data segments. Specifically, the same feature extraction, quantization, and agreement procedures can be reapplied over consecutive time windows without requiring a reset or re-synchronization. This allows for continuous or on-demand key refreshment, which we refer to as “real-time” in the sense that the key material adapts to the ongoing biometric input and can be updated incrementally as part of normal system operation.}

While our scheme demonstrates significant potential for enhancing mHealth security, we acknowledge certain limitations in our current research phase. Due to equipment constraints, our method has not yet been deployed on actual medical devices. Transitioning from theoretical models to real-world applications involves complex challenges, including the significant costs of software development and adaptation for various medical platforms.
The variability of physiological signals also introduces a challenge in the practical application of these encryption techniques.
Factors such as stress, physical activity, and varying health conditions can lead to fluctuations in the physiological parameters used to generate encryption keys, potentially compromising their reliability and security. For instance, a patient's heart rate may vary significantly during physical exertion or emotional stress, altering the characteristics of the signal used for key generation. This variability necessitates the development of sophisticated algorithms capable of normalizing these fluctuations to ensure that the encryption keys remain consistent and secure over time.
Moreover, ethical considerations, particularly those related to user safety, privacy, and data protection, necessitate careful consideration and compliance with institutional protocols \cite{shu2026trustworthy}. These factors underscore the need for further research and development to bridge the gap between simulation-based validation and practical implementation in medical settings. Despite these limitations, our simulations offer valuable insights, laying a foundation for future advancements in secure mHealth communications.

\section{Conclusion}
\label{conclusion}
The rapid development of mobile health technologies has created an increasing demand for secure communication mechanisms to protect sensitive health data during transmission. In this paper, we introduced SigK, a novel key agreement scheme that leverages rPPG signals to enable secure communication among contactless medical devices. Experimental results on three publicly available datasets demonstrate the feasibility and effectiveness of the proposed scheme. In addition, the comprehensive formal security proof further validates its capability to protect the privacy and security of mobile health data transmission. Future work will focus on implementing SigK in real-world medical scenarios and evaluating its adaptability to diverse mHealth devices and application environments, thereby further demonstrating its practical value and potential for broader deployment.

\ifCLASSOPTIONcaptionsoff
  \newpage
\fi

\bibliographystyle{IEEEtran}
\bibliography{Bibliography}

\vfill

\end{document}